\documentclass{article}%
\usepackage{amsmath}
\usepackage{amsfonts}
\usepackage{amssymb}
\usepackage{graphicx}%
\providecommand{\U}[1]{\protect\rule{.1in}{.1in}}
\newtheorem{theorem}{Theorem}

\newtheorem{remark}[theorem]{Remark}

\begin{document}

\title{The Dirac-Hestenes Equation and its Relation with the Relativistic de
Broglie-Bohm Theory}
\author{A. M. Moya$^{(\ast)}$, W. A. Rodrigues Jr.$^{(\ast\ast)}$ and S. A.
Wainer$^{(\ast\ast)}$
\and $^{(\ast)}$National Technological University-Haedo Regional Faculty
\and anmanumoya@gmail.com
\and Paris 532, 1706 Haedo, Buenos Aires, Argentina
\and $^{(\ast\ast)}$Institute of Mathematics, Statistics and Scientific Computation
\and IMECC-UNICAMP
\and walrod@ime.unicamp.br \ samuel wainer@ime.unicamp.br}
\date{October 27 2016}
\maketitle

\begin{abstract}
In this paper we provide\ using the Clifford and spin-Clifford formalism and
some few results of the extensor calculus a derivation of the conservation
laws that follow directly from the Dirac-Hestenes equation (DHE) describing a
Dirac-Hestenes spinor field (DHSF) in interaction with an external
electromagnetic field without using the Lagrangian formalism. In particular,
we show that the energy-momentum and total angular momentum extensors of a
DHSF is not conserved in spacetime regions permitting the existence of a null
electromagnetic field $F$ but a non null electromagnetic potential $\emph{A}$.
These results have been used together with some others recently obtained
(e.g., that the classical \ relativistic Hamilton-Jacobi equation is
equivalent to a DHE satisfied by a particular class of DHSF) to obtain the
correct relativistic quantum potential when the Dirac theory is interpreted as
a de Broglie-Bohm theory. Some results appearing in the literature on this
issue are criticized and the origin of some misconceptions is detailed with a
rigorous mathematical analysis.

\end{abstract}

\section{Introduction}

Using the Clifford and spin Clifford bundle formalisms and some few results of
the theory of extensor calculus we derive the conservation laws for the
probability current, energy-momentum and total angular momentum
\emph{extensor} \emph{fields} resulting from the structure of Dirac-Hestenes
\ equation (DHE) satisfied by a Dirac-Hestenes spinor field (DHSF) in
interaction with an electromagnetic field \ without using the Lagrangian
formalism\footnote{We recall that the concepts of multiform functions and
extensor fields permit us to give a nice Lagrangian (and Hamiltonian)
formalism for all relativistic fields and in particular to derive the
conserved energy-momentum, spin and angular momentum extensor fields for
interacting relativistic fields. Preliminary details of our formalism has been
first presented in \cite{rsv1993,m1999,mfr2001,} and further elaborated in
\cite{rc2007}.}. We show in a quite simple way that the energy-momentum
extensor field\footnote{Related to theTetrode tensor of the Dirac spinor
field.} of the DHSF is not symmetric and that its antisymmetric part is the
\emph{source }of the \emph{spin extensor field} of the free DHSF. Moreover we
show that the conservation laws for the energy-momentum and total angular
momentum extensor for the coupled system of the DHSF and the electromagnetic
field is very different from what is expected from a classical point of view.
In particular these laws show in an elegant way that in spacetime regions with
topology permitting the existence of an electromagnetic field potential
$A\neq0$ but with $F=0$ the energy-momentum and total angular momentum of the
sole DHSF are not conserved. This is clearly according to our view the origin
of the Bohm-Aharonov effect.

In Section 3 we briefly recall results obtained in \cite{rw2016a} where it was
shown that (a): that the classical\ relativistic Hamilton-Jacobi equation
(HJE) \ for a charged electrical particle in interaction with the
electromagnetic field is equivalent to a DHE satisfied by a special class of
DHSF that we call \emph{classical }DHSF which are characterized by having the
Takabayashi angle equal to $0$ or $\pi$ and moreover that the DHE satisfied by
a classical DHSF is equivalent to the HJE; (b) The identification of the
correct relativistic quantum potential resulting from the DHE in a de
Broglie-Bohm approach leading to a HJE like equation for the motion of
(spinning) charged particle.

Equipped with these results and the ones of Section 2 we analyze in Section 3
results of \cite{hc1,hc2} were authors though that they have disclosed the
relativistic quantum potential for the Dirac particle. We prove that the
results of those papers are equivocated by explicitly showing with detailed
calculations were authors get mislead.

In Section 4 we present our conclusions and in the Appendix we recall the
notation and some results of the Clifford and spin-Clifford bundles formalism
used in the paper.

\section{Dirac-Hestenes Equation (DHE) and Conserved Currents}

With the notations introduced in the Appendix the DHE for $\phi_{L}$ $\in
\sec\mathcal{C\ell}^{0}(M,\mathtt{\eta})$ \ and $\phi_{R}$ $\in\sec
\mathcal{C\ell}^{0}(M,\mathtt{\eta})$ \ in the presence of an electromagnetic
potential $A\in%
{\textstyle\bigwedge\nolimits^{1}}
T^{\ast}M$ $\hookrightarrow\sec\mathcal{C\ell}^{0}(M,\mathtt{\eta})$ are
\begin{align}
\boldsymbol{\partial}\phi_{L}\gamma_{21}-eA\phi_{L}-m\phi_{L}\gamma_{0}  &
=0,\label{1}\\
\gamma_{12}\phi_{R}\overleftarrow{\boldsymbol{\partial}}-e\phi_{R}%
A-m\gamma_{0}\phi_{R}  &  =0 \label{1A}%
\end{align}
Multiplying Eq.(\ref{1}) on the right by $\gamma_{0}$ $\phi_{R}$we get%
\begin{equation}
\boldsymbol{\partial}\phi_{L}\gamma_{012}\phi_{R}-eA\phi_{L}\gamma_{0}\phi
_{R}-m\phi_{L}\phi_{R}=0 \label{2}%
\end{equation}

Now, taking notice that for any odd $O\in\sec$ ($%
{\textstyle\bigwedge\nolimits^{1}}
T^{\ast}M+%
{\textstyle\bigwedge\nolimits^{3}}
T^{\ast}M$)$\hookrightarrow\sec\mathcal{C\ell}(M,\mathtt{\eta})$ it is%
\begin{equation}
\langle O\rangle_{1}=\frac{1}{2}(O+\tilde{O}) \label{3}%
\end{equation}
and that%
\begin{equation}
\partial_{\mu}(\phi_{L}\gamma_{012}\phi_{R})=\partial_{\mu}\phi_{L}%
\gamma_{012}\phi_{R}+\phi_{L}\gamma_{012}\partial_{\mu}\phi_{R}, \label{4}%
\end{equation}
we have%
\begin{equation}
\langle\partial_{\mu}\phi_{L}\gamma_{012}\phi_{R}\rangle_{1}=\frac{1}%
{2}\left(  \partial_{\mu}\phi_{L}\gamma_{012}\phi_{R}-\phi_{L}\gamma
_{012}\partial_{\mu}\phi_{R}\right)  . \label{5}%
\end{equation}
Summing Eq.(\ref{4}) (multiplied by $1/2)$ with Eq.(\ref{5}) and multiplying
on the left by $\gamma^{\mu}$ we get
\begin{equation}
\frac{1}{2}\boldsymbol{\partial}(\phi_{L}\gamma_{012}\phi_{R})+\gamma^{\mu
}\langle\partial_{\mu}\phi_{L}\gamma_{012}\phi_{R}\rangle_{1}%
=\boldsymbol{\partial}\phi_{L}\gamma_{012}\phi_{R}. \label{6}%
\end{equation}
Next, using Eq.(\ref{1}) on the right side of Eq.(\ref{3}) we get%
\begin{equation}
\frac{1}{2}\boldsymbol{\partial}(\gamma_{5}s)+\gamma^{\mu}\langle\partial
_{\mu}\phi_{L}\gamma_{012}\phi_{R}\rangle_{1}=eAJ+m\phi_{L}\phi_{R} \label{7}%
\end{equation}
where
\begin{equation}
s=\phi_{L}\gamma^{3}\phi_{R}\text{ ~and ~}J=\phi_{L}\gamma^{0}\phi_{R}
\label{8}%
\end{equation}
are bilinear invariants of Dirac theory called respectively the spin $1$-form
field and the probability current $1$-form field.\footnote{Details in
\cite{rc2007}.}

\subsection{The Conserved Probability Current}

From Eq.(\ref{1}) we also have%
\begin{equation}
\boldsymbol{\partial}\phi_{L}\gamma_{0}\phi_{R}=eA\phi_{L}\gamma_{012}\phi
_{R}+m\phi_{L}\gamma_{21} \label{1c}%
\end{equation}

Now, taking also into account that for $O\in\sec$ ($%
{\textstyle\bigwedge\nolimits^{1}}
T^{\ast}M+%
{\textstyle\bigwedge\nolimits^{3}}
T^{\ast}M$)$\hookrightarrow\sec\mathcal{C\ell}(M,\mathtt{\eta})$ it is
\begin{equation}
\langle O\rangle_{3}=\frac{1}{2}(O-\tilde{O}) \label{9}%
\end{equation}
we also have the identities%
\begin{align}
\partial_{\mu}(\phi_{L}\gamma_{0}\phi_{R})  &  =\partial_{\mu}\phi_{L}%
\gamma_{0}\phi_{R}+\phi_{L}\gamma_{0}\partial_{\mu}\phi_{R},\label{10}\\
\langle\partial_{\mu}\phi_{L}\gamma_{0}\phi_{R}\rangle_{3}  &  =\frac{1}%
{2}\left(  \partial_{\mu}\phi_{L}\gamma_{0}\phi_{R}-\phi_{L}\gamma_{0}%
\partial_{\mu}\phi_{R}\right)  \label{10a}%
\end{align}
and summing Eqs.(\ref{10}) and (\ref{10a}) and multiplying on the left by
$\gamma^{\mu}$ we get%
\begin{equation}
\frac{1}{2}\boldsymbol{\partial}(\phi_{L}\gamma_{0}\phi_{R})+\gamma^{\mu
}\langle\partial_{\mu}\phi_{L}\gamma_{0}\phi_{R}\rangle_{3}%
=\boldsymbol{\partial}\phi_{L}\gamma_{0}\phi_{R} \label{11}%
\end{equation}
Using Eq.(\ref{1c}) in Eq.(\ref{11}) we get \
\begin{equation}
\boldsymbol{\partial}J+\gamma^{\mu}\langle\partial_{\mu}\phi_{L}\gamma_{0}%
\phi_{R}\rangle_{3}=-eA\gamma_{5}s. \label{12}%
\end{equation}
Now, from Eq.(\ref{12}) we get
\begin{align}
\langle\boldsymbol{\partial}J+\gamma^{\mu}\langle\partial_{\mu}\phi_{L}%
\gamma_{0}\phi_{R}\rangle_{3}\rangle_{0}  &  =\langle\boldsymbol{\partial
}J\rangle_{0}=\boldsymbol{\partial\lrcorner}J,\nonumber\\
\langle-eA\gamma_{5}s\rangle_{0}  &  =-e\langle\langle A\gamma_{5}s\rangle
_{2}+\langle A\gamma_{5}s\rangle_{2}\rangle_{0}=0 \label{13}%
\end{align}
and finally we have the conservation law for the probability
current\footnote{Recall that $\boldsymbol{\partial}=d-\delta$, where
$d=\boldsymbol{\partial\wedge}$ is the differential operator and
$\delta=-\boldsymbol{\partial}\boldsymbol{\lrcorner}$ is Hodge codifferential
operator}, i.e.,
\begin{equation}
\boldsymbol{\partial\lrcorner}J=-\delta J=0. \label{14}%
\end{equation}

\section{The Energy-Momentum Extensor of the Dirac-Hestenes Field}

Recalling that for $v\in\sec$ $%
{\textstyle\bigwedge\nolimits^{1}}
T^{\ast}M\hookrightarrow\sec\mathcal{C\ell}(M,\mathtt{\eta})$ and
$\mathcal{C\in}\sec\mathcal{C\ell}(M,\mathtt{\eta})$ the following identity
holds \cite{rc2007}
\begin{equation}
\boldsymbol{\partial}(v\mathcal{C})=\boldsymbol{\partial}(v)\mathcal{C-}%
v(\boldsymbol{\partial}\mathcal{C})+2(v\lrcorner\boldsymbol{\partial
)}\mathcal{C} \label{16}%
\end{equation}
we immediately get from Eq.(\ref{1}) after some trivial algebra that
\begin{equation}
\boldsymbol{\partial}^{2}\phi_{L}=(e^{2}A^{2}-m^{2})\phi_{L}%
-e[(\boldsymbol{\partial}A)\phi_{L}+2(A\lrcorner\boldsymbol{\partial)}\phi
_{L}]\gamma_{21}. \label{15}%
\end{equation}
Next, recalling again the identity given by Eq.(\ref{3}) we have
\begin{equation}
\langle\boldsymbol{\partial}^{2}\phi_{L}\gamma_{012}\phi_{R}\rangle_{1}%
=\frac{1}{2}\left(  \boldsymbol{\partial}^{2}\phi_{L}\gamma_{012}\phi_{R}%
-\phi_{L}\gamma_{012}\boldsymbol{\partial}^{2}\phi_{R}\right)  . \label{17}%
\end{equation}
Next, using Eq.(\ref{17}) on the right side of Eq.(\ref{15}) multiplied by
$\gamma_{012}\phi_{R}$\ we get%
\begin{equation}
\langle\boldsymbol{\partial}^{2}\phi_{L}\gamma_{012}\phi_{R}\rangle
_{1}=eF\llcorner J+e[(\boldsymbol{\partial\lrcorner}A)J+(A\lrcorner
\boldsymbol{\partial})J] \label{18}%
\end{equation}
with $F=\boldsymbol{\partial}\wedge A=dA\in\sec$ $%
{\textstyle\bigwedge\nolimits^{2}}
T^{\ast}M\hookrightarrow\sec\mathcal{C\ell}(M,\mathtt{\eta})$ is the Faraday field.

We now write Eq.(\ref{18}) in a convenient form which permit us to introduce
the energy-momentum $(1,1)$-extensor field for the Dirac-Hestenes field, i.e.,
the object%
\begin{align}
\boldsymbol{T}  &  :\sec%
{\textstyle\bigwedge\nolimits^{1}}
T^{\ast}M\hookrightarrow\sec\mathcal{C\ell}(M,\mathtt{\eta})\rightarrow\sec%
{\textstyle\bigwedge\nolimits^{1}}
T^{\ast}M\hookrightarrow\sec\mathcal{C\ell}(M,\mathtt{\eta}),\nonumber\\
n  &  \mapsto\boldsymbol{T}(n) \label{19}%
\end{align}
such that
\begin{equation}
\boldsymbol{T}(\gamma^{\mu})\cdot\gamma^{\nu}=T^{\mu\nu} \label{20}%
\end{equation}
are the components of the energy-momentum tensor of the Dirac field. In order
to proceed we introduce the differential operators
\begin{equation}
\boldsymbol{\partial}_{n}:=\gamma^{\mu}\frac{\partial}{\partial n^{\mu}}\text{
~and ~}\boldsymbol{\partial}_{n}\cdot\boldsymbol{\partial}=\eta^{\mu\nu}%
\frac{\partial}{\partial n^{\mu}}\frac{\partial}{\partial\mathrm{x}^{\nu}}
\label{21}%
\end{equation}
acting on the bundle of the $(1,1)$-extensor fields. Recalling moreover that
for any $\mathcal{C\in}\sec\mathcal{C\ell}(M,\mathtt{\eta})$ it is%
\begin{equation}
\boldsymbol{\partial}_{n}\cdot\boldsymbol{\partial(}n\cdot\partial
\mathcal{C)}=\boldsymbol{\partial}^{2}\mathcal{C} \label{22}%
\end{equation}
and
\begin{equation}
\partial^{\mu}\mathcal{C}=\gamma^{\mu}\cdot\boldsymbol{\partial}%
_{n}\boldsymbol{(}n\cdot\partial\mathcal{C)} \label{23}%
\end{equation}
the left side of Eq.(\ref{18}) can be written as%
\begin{equation}
\langle\boldsymbol{\partial}^{2}\phi_{L}\gamma_{012}\phi_{R}\rangle
_{1}=\boldsymbol{\partial}_{n}\cdot\boldsymbol{\partial}\langle(n\cdot
\boldsymbol{\partial)}\phi_{L}\gamma_{012}\phi_{R}\rangle_{1}. \label{24}%
\end{equation}
Now, observe that it is
\begin{align}
(\boldsymbol{\partial\lrcorner}A)J+(A\lrcorner\boldsymbol{\partial})J  &
=\gamma^{\mu}\cdot(\boldsymbol{\partial}_{\mu}A)J+\gamma^{\mu}\cdot
A\boldsymbol{\partial}_{\mu}J\nonumber\\
&  =\gamma^{\mu}\cdot\boldsymbol{\partial}_{n}[(\partial_{\mu}n)\cdot
AJ+n\cdot(\partial_{\mu}A)J+n\cdot A\partial_{\mu}J]\nonumber\\
&  =\gamma^{\mu}\cdot\boldsymbol{\partial}_{n}\partial_{\mu}(n\cdot
AJ)\nonumber\\
&  =\boldsymbol{\partial}_{n}\cdot\boldsymbol{\partial(}n\cdot
AJ\boldsymbol{).} \label{25}%
\end{align}
Then, using Eqs.(\ref{24}) and (\ref{25}) we can write Eq.(\ref{18})
as\footnote{Recall that classicaly $eF\lrcorner J$ represents the Lorentz
force law.}%
\begin{equation}
\boldsymbol{\partial}_{n}\cdot\boldsymbol{\partial\lbrack}\langle
(n\cdot\boldsymbol{\partial)}\phi_{L}\gamma_{012}\phi_{R}\rangle_{1}-en\cdot
AJ]=eF\lrcorner J. \label{26}%
\end{equation}
We now write
\begin{equation}
\boldsymbol{T}^{\dagger}(n):=\langle(n\cdot\boldsymbol{\partial)}\phi
_{L}\gamma_{012}\phi_{R}\rangle_{1}-en\cdot AJ \label{28}%
\end{equation}
where $\boldsymbol{T}^{\dagger}$ denotes the adjoint of a $(1,1)$-extensor
field $\boldsymbol{T}$, i.e., for $n,m\in\sec%
{\textstyle\bigwedge\nolimits^{1}}
T^{\ast}M\hookrightarrow\sec\mathcal{C\ell}(M,\mathtt{\eta})$ it is
\begin{equation}
\boldsymbol{T}(n)\cdot m=n\cdot\boldsymbol{T}^{\dagger}(m) \label{27}%
\end{equation}
and write Eq.(\ref{26}) as%
\begin{equation}
\boldsymbol{\partial}_{n}\cdot\boldsymbol{\partial T}^{\dagger}(n)=eF\lrcorner
J \label{29}%
\end{equation}
which can also be written as%
\[
\boldsymbol{\partial\cdot T}^{\dagger}(\gamma_{\mu})=e(F\lrcorner
J)\cdot\gamma_{\mu}%
\]

When $A=0$ we write $\boldsymbol{T}^{\dagger}(n)=\boldsymbol{T}_{D}^{\dagger
}(n)$
\begin{equation}
\boldsymbol{T}_{D}^{\dagger}(n)=\langle(n\cdot\boldsymbol{\partial)}\phi
_{L}\gamma_{012}\phi_{R}\rangle_{1} \label{30}%
\end{equation}
which is the energy-momentum extensor field of the Dirac-Hestenes field.
Obviously, in a region where $\ F=0$ and $A=0$, $\boldsymbol{T}_{D}^{\dagger
}(n)$ is conserved, i.e.,
\begin{equation}
\boldsymbol{\partial}_{n}\cdot\boldsymbol{\partial T}_{D}^{\dagger}(n)=0.
\label{31}%
\end{equation}
We write
\begin{equation}
\boldsymbol{T}^{\mu}=\boldsymbol{T}(\gamma^{\mu}):=T^{\mu\nu}\gamma_{\nu}
\label{32}%
\end{equation}
and of course%
\begin{equation}
\boldsymbol{T}^{\dagger\mu}=\boldsymbol{T}^{\dagger}(\gamma^{\mu})=T^{\nu\mu
}\gamma_{\nu}. \label{33}%
\end{equation}
The objects $\boldsymbol{T}^{\mu}\in\sec%
{\textstyle\bigwedge\nolimits^{1}}
T^{\ast}M\hookrightarrow\sec\mathcal{C\ell}(M,\mathtt{\eta})$are said to be
the energy-momentum 1-form fields of the Dirac field in interaction with an
electromagnetic field. The $T^{\mu\nu}$ are the components of the so-called
Tetrode energy-momentum tensor of the Dirac field and as it is well known and
will be shown below it is%
\begin{equation}
T^{\mu\nu}\neq T^{\nu\mu}, \label{34}%
\end{equation}
i.e., the Tetrode tensor is not symmetric, something that has a nontrivial
implication shown in subsection \ref{sourcesp}, namely that the antisymmetric
part of the energy-momentum tensor is the source of the spin field of the
Dirac field.

\subsection{The Trace and the Bif of the Extensor field $\boldsymbol{T}$}

To proceed, we need some results, in particular the \emph{trace} \ and the
\emph{bif} of the extensor field $\boldsymbol{T}$ defined \cite{rc2007}
respectively by by%
\begin{align}
\mathrm{tr}(\boldsymbol{T})  &  =\mathrm{tr}(\boldsymbol{T}^{\dagger}%
):=\gamma^{\mu}\cdot\boldsymbol{T}^{\dagger}(\gamma_{\mu})=\boldsymbol{T}%
^{\dagger}(\gamma^{\mu})\cdot\gamma_{\mu}.\label{35}\\
\mathrm{bif}(\boldsymbol{T})  &  =-\mathrm{bif}(\boldsymbol{T}^{\dagger
})=-\boldsymbol{T}^{\dagger}(\gamma_{\mu})\wedge\gamma^{\mu}=\gamma^{\mu
}\wedge\boldsymbol{T}^{\dagger}(\gamma_{\mu}). \label{35a}%
\end{align}
Now, the $\langle\rangle_{0}$ part of Eq.(\ref{7}) is taking into account
that\footnote{Note that $\phi_{L}\cdot\phi_{R}=\phi_{L}\cdot\phi_{L}=\phi
_{R}\cdot\phi_{R}.$}
\begin{equation}
\gamma^{\mu}\cdot\langle\partial_{\mu}\phi_{L}\gamma_{012}\phi_{R}\rangle
_{1}=eA\cdot J+m\phi_{L}\cdot\phi_{R}. \label{36}%
\end{equation}
Then,%
\begin{equation}
tr(\boldsymbol{T})=m\phi_{L}\cdot\phi_{R} \label{37}%
\end{equation}

Now, taking the $\langle\rangle_{0}$ part of Eq.(\ref{7}) we get
\begin{equation}
\frac{1}{2}\boldsymbol{\partial\lrcorner}(\gamma_{5}s)+\gamma^{\mu}%
\wedge\langle\partial_{\mu}\phi_{L}\gamma_{012}\phi_{R}\rangle_{1}=eA\wedge J
\label{38}%
\end{equation}
and then recalling also that $\boldsymbol{\partial\lrcorner}(\gamma
_{5}s)=-\gamma_{5}\boldsymbol{\partial\wedge}s$%
\begin{equation}
\mathrm{bif}(\boldsymbol{T})=\frac{1}{2}\gamma_{5}\boldsymbol{\partial\wedge
}s. \label{39}%
\end{equation}

Also, since
\begin{equation}
\boldsymbol{T}^{\dagger}(n)-\boldsymbol{T}(n)=n\lrcorner\mathrm{bif}%
(\boldsymbol{T}) \label{40}%
\end{equation}
we have
\begin{gather}
\boldsymbol{\partial}_{n}\cdot\boldsymbol{\partial T}^{\dagger}%
(n)-\boldsymbol{\partial}_{n}\cdot\boldsymbol{\partial T}%
(n)=\boldsymbol{\partial}_{n}\cdot\boldsymbol{\partial}(n\lrcorner
\mathrm{bif}(\boldsymbol{T}))\nonumber\\
=\gamma^{\mu}\cdot\boldsymbol{\partial}_{n}[\partial_{\mu}n\lrcorner
\mathrm{bif}(\boldsymbol{T})+n\lrcorner\partial_{\mu}\mathrm{bif}%
(\boldsymbol{T})]\nonumber\\
=\boldsymbol{\partial}\lrcorner\mathrm{bif}(\boldsymbol{T}%
)=-\boldsymbol{\partial}\lrcorner(\boldsymbol{\partial\lrcorner}(\gamma
_{5}s)=-\boldsymbol{\delta}\lrcorner(\boldsymbol{\delta\lrcorner}(\gamma
_{5}s)=0 \label{41a}%
\end{gather}
which means that
\begin{equation}
\boldsymbol{\partial}_{n}\cdot\boldsymbol{\partial T}^{\dagger}%
(n)=\boldsymbol{\partial}_{n}\cdot\boldsymbol{\partial T}(n). \label{42}%
\end{equation}

\subsection{The Source of Spin Extensor of the Dirac-Hestenes
Field\label{sourcesp}}

Taking advantage that Minkowski spacetime is parallelizable we introduce the
$1$-form
\begin{equation}
\boldsymbol{x}:=x^{\mu}\gamma_{\mu} \label{43}%
\end{equation}
which will be called the position $1$-form of a point of $x\in M$ relative to
the point $x_{0}\in M$ which is the one mapped to the origin of the
coordinates in $\mathbb{R}^{4}$. If we make the exterior product of both
members of Eq.(\ref{29}) by $\boldsymbol{x}$ we get%
\begin{equation}
\lbrack\boldsymbol{\partial}_{n}\cdot\boldsymbol{\partial T}(n)]\wedge
\boldsymbol{x}=e(F\llcorner J)\wedge\boldsymbol{x} \label{44}%
\end{equation}
using the identity%
\begin{equation}
\boldsymbol{\partial}_{n}\cdot\boldsymbol{\partial\lbrack\boldsymbol{T}%
}(n)\boldsymbol{\wedge\boldsymbol{x}]}=[\boldsymbol{\partial}_{n}%
\cdot\boldsymbol{\partial T}(n)]\wedge\boldsymbol{x}+\mathrm{bif}%
(\boldsymbol{T}) \label{45}%
\end{equation}
in Eq.(\ref{44}) we get%
\begin{equation}
\boldsymbol{\partial}_{n}\cdot\boldsymbol{\partial\lbrack\boldsymbol{T}%
}(n)\boldsymbol{\wedge\boldsymbol{x}]}-\mathrm{bif}(\boldsymbol{T}%
)=e(F\llcorner J)\wedge\boldsymbol{x}. \label{46}%
\end{equation}

Now, we have that
\begin{gather*}
\mathrm{bif}(\boldsymbol{T})=\frac{1}{2}\gamma_{5}\boldsymbol{\partial\wedge
}s=-\frac{1}{2}\gamma_{5}\boldsymbol{\partial}_{\mu}(s\wedge\gamma^{\mu})\\
=-\frac{1}{2}\gamma^{\mu}\cdot\boldsymbol{\partial}_{n}\gamma_{5}%
[\boldsymbol{\partial}_{\mu}s\wedge n+s\wedge\boldsymbol{\partial}_{\mu
}\boldsymbol{x}]\\
-\gamma^{\mu}\cdot\boldsymbol{\partial}_{n}\boldsymbol{\partial}_{\mu}\frac
{1}{2}\gamma_{5}(s\wedge n)=-\boldsymbol{\partial}_{n}\cdot
\boldsymbol{\partial}\gamma_{5}(s\wedge n)
\end{gather*}
and introducing the $(1,2)$-extensor field
\begin{gather}
\boldsymbol{S}:\sec%
{\textstyle\bigwedge\nolimits^{1}}
T^{\ast}M\hookrightarrow\sec\mathcal{C\ell}(M,\mathtt{\eta})\rightarrow\sec%
{\textstyle\bigwedge\nolimits^{2}}
T^{\ast}M\hookrightarrow\sec\mathcal{C\ell}(M,\mathtt{\eta}),\nonumber\\
\boldsymbol{S}(n):=\frac{1}{2}\gamma_{5}(s\wedge n) \label{47}%
\end{gather}
called the \emph{spin extensor field} of the Dirac field we have that
\begin{equation}
\mathrm{bif}(\boldsymbol{T})=-\boldsymbol{\partial}_{n}\cdot
\boldsymbol{\partial S}(n) \label{48}%
\end{equation}
If we contract Eq.(\ref{48}) on the left taking account of Eq.(\ref{40}) we
get the fundamental result%
\begin{equation}
\boldsymbol{T}(n)-\boldsymbol{T}^{\dagger}(n)=n\lrcorner(\boldsymbol{\partial
}_{n}\cdot\boldsymbol{\partial S}(n)) \label{49}%
\end{equation}
which says that the source of the spin extensor is the antisymmetric part of
the energy-momentum extensor. From Eq.(\ref{49}) we can immediately get that
\begin{align}
T^{\mu\nu}-T^{\nu\mu}  &  =-\gamma^{\nu}\lrcorner(\gamma^{\mu}\lrcorner
(\boldsymbol{\partial}_{\kappa}\boldsymbol{S}(\gamma^{\kappa})))\nonumber\\
&  =-(\gamma^{\nu}\wedge\gamma^{\mu})\lrcorner\boldsymbol{\partial}_{\kappa
}\boldsymbol{S}(\gamma^{\kappa}). \label{50}%
\end{align}

Finally calling
\begin{equation}
\boldsymbol{J}(n):=\boldsymbol{T}(n)\wedge\boldsymbol{x}+\boldsymbol{S}(n)
\label{51}%
\end{equation}
t we get using Eq.(\ref{48}) on the left side of Eq.(\ref{46})
\begin{equation}
\boldsymbol{\partial}_{n}\cdot\boldsymbol{\partial J}(n)=e(F\llcorner
J)\wedge\boldsymbol{x} \label{52}%
\end{equation}
Recalling that $(F\llcorner J)\wedge\boldsymbol{x}$ is the torque produced by
the Lorentz force $e(F\llcorner J)$ we see that Eq.(\ref{52}) says explicitly
that in the presence of an external electromagnetic field the total angular
momentum of the Dirac field is not conserved.

\subsection{Origin of the Bohm-Aharonov Effect}%

\begin{equation}
\boldsymbol{\partial}_{n}\cdot\boldsymbol{\partial T}^{\dagger}(n)=eF\lrcorner
J
\end{equation}

In our opinion Eq.(\ref{29}) and Eq.(\ref{52}) encode the origin of the
Bohm-Aharonov effect. Indeed, recalling Eq.(\ref{28}) and Eq.(\ref{51}) we
have%
\begin{align*}
\boldsymbol{T}(n)  &  =\boldsymbol{T}_{D}(n)-en\cdot AJ,\\
\boldsymbol{J}(n)  &  :=\boldsymbol{T}_{D}(n)\wedge\boldsymbol{x}%
+\boldsymbol{S}(n)-en\cdot AJ\wedge\boldsymbol{x}=\boldsymbol{J}%
_{D}(n)-en\cdot AJ\wedge\boldsymbol{x}%
\end{align*}
So, it follows that even in spacetime regions with topology permitting that
$F=0$ but $A\neq0$ we must have (with $A=A^{\mu}\gamma_{\mu}$) from Eqs.
(\ref{29}) and (\ref{52})%
\begin{align}
\boldsymbol{\partial}_{n}\cdot\boldsymbol{\partial T}_{D}(n)  &
=\partial_{\mu}\boldsymbol{T}_{D}(\gamma^{\mu})=e\boldsymbol{\partial}%
_{n}\cdot\boldsymbol{\partial(}n\cdot AJ)=\partial_{\mu}(A^{\mu}J)\label{53}\\
\boldsymbol{\partial}_{n}\cdot\boldsymbol{\partial J}_{D}(n)  &
=\partial_{\mu}\boldsymbol{J}_{D}(\gamma^{\mu})=\boldsymbol{\partial}_{n}%
\cdot\boldsymbol{\partial}[n\cdot AJ\wedge\boldsymbol{x}]=\partial_{\mu
}(A^{\mu}J\wedge\boldsymbol{x}) \label{54}%
\end{align}
which says that in regions where $F=0$ but $A\neq0$ the energy-momentum
$\boldsymbol{T}_{D}(n)$ and the total angular-momentum extensor
$\boldsymbol{J}_{D}(n)$ of the Dirac field are \emph{not }conserved. This
certainly means that a particle described by a Dirac field entering a region
with topology permitting that $F=0$ but $A\neq0$ will follows paths distinct
from the ones where it enter a region where $F=0$ and $A=0$. This is exactly
what is predicted in the Bohm-Aharonov effect and the experimental study of
the possible trajectories (as done, e.g., in) in the de Broglie-Bohm
interpretation of the Dirac equation seems a good test for such interpretation
of quantum mechanics.

\section{The Correct Relativistic Quantum Potential}

In this section we analyze in details papers \cite{hc1,hc2} showing that it
contains erroneous results and thus an erroneous identification of the quantum
potential. Moreover, we recall form our recent paper \cite{rw2016a} where we
derive the correct generalized Hamilton-Jacobi equation which follows from the
DHE satisfied by a DHSF and identifies the correct relativistic quantum potential.

So, to start our analysis, recall that from Eq.(\ref{30}) we can write for the
energy-momentum $1$-form fields $\boldsymbol{T}_{D}^{\dagger^{\mu}}$ that
\begin{equation}
\boldsymbol{T}_{D}^{\dagger^{\mu}}=\langle\partial^{\mu}\phi_{L}\gamma
_{012}\phi_{R}\rangle_{1}=\frac{1}{2}\left(  \partial^{\mu}\phi_{L}%
\gamma_{012}\phi_{R}-\phi_{L}\gamma_{012}\partial^{\mu}\phi_{R}\right)
\label{55}%
\end{equation}

In \cite{hc1,hc2} authors define the objects
\begin{equation}
\rho\mathbf{P}^{\mu}=\frac{1}{2}\left(  \partial^{\mu}\phi_{L}\gamma_{012}%
\phi_{R}-\phi_{L}\gamma_{012}\partial^{\mu}\phi_{R}\right)  \label{56}%
\end{equation}%
\begin{gather}
\rho\mathbf{W}^{\mu}=-\frac{1}{2}\partial^{\mu}(\phi_{L}\gamma_{012}\phi
_{R})\nonumber\\
=-\frac{1}{2}[\partial^{\mu}\phi_{L}\gamma_{012}\phi_{R}+\phi_{L}\gamma
_{012}\partial^{\mu}\phi_{R}]=-\langle\partial^{\mu}\phi_{L}\gamma_{012}%
\phi_{R}\rangle_{3} \label{57}%
\end{gather}
which taking into account Eqs.(\ref{3}) and (\ref{9}) permit us to write%
\begin{align}
\rho\mathbf{P}^{\mu}  &  =\boldsymbol{T}_{D}^{\dagger^{\mu}}=\langle
\partial^{\mu}\phi_{L}\gamma_{012}\phi_{R}\rangle_{1}\in\sec%
{\textstyle\bigwedge\nolimits^{1}}
T^{\ast}M\hookrightarrow\sec\mathcal{C\ell}(M,\mathtt{\eta})\label{58}\\
\rho\mathbf{W}^{\mu}  &  =-\langle\partial^{\mu}\phi_{L}\gamma_{012}\phi
_{R}\rangle_{3}\in\sec%
{\textstyle\bigwedge\nolimits^{3}}
T^{\ast}M\hookrightarrow\sec\mathcal{C\ell}(M,\mathtt{\eta}) \label{59}%
\end{align}
Authors of \cite{hc1,hc2} also introduce the object\footnote{Take notice that
the object $\mathbf{J}$ is not the same object we call $J$ (the probability
current) given by Eq.(\ref{8}) and is also different from the total angular
momentum extensor which we denoted by $\boldsymbol{J}(n)$.}
\begin{equation}
\rho\mathbf{J}:=\phi_{L}\gamma_{012}\phi_{R}\in\sec%
{\textstyle\bigwedge\nolimits^{1}}
T^{\ast}M\hookrightarrow\sec\mathcal{C\ell}(M,\mathtt{\eta}) \label{60}%
\end{equation}

Now, recall that for a Dirac particle moving in a region where $A=0$,
Eq.(\ref{15}) becomes%

\begin{equation}
\boldsymbol{\partial}^{2}\phi_{L}+m^{2}\phi_{L}=0 \label{61}%
\end{equation}
and of course we have also that
\begin{equation}
\boldsymbol{\partial}^{2}\phi_{R}+m^{2}\phi_{R}=0. \label{62}%
\end{equation}
Multiplying Eq.(\ref{61}) on the right by $\phi_{R}$ and Eq.(\ref{62}) on the
right by $\phi_{L}$ and taking first the sum of these equations and second
their difference we get $\partial_{\mu}\partial^{\mu}$%
\begin{equation}
\partial_{\mu}(\partial^{\mu}\phi_{L})\phi_{R}+\phi_{L}(\partial_{\mu}%
\partial^{\mu}\phi_{R})+2m^{2}\phi_{L}\phi_{R}=0 \label{63}%
\end{equation}
and
\begin{equation}
\phi_{L}(\partial_{\mu}\partial^{\mu}\phi_{R})-\partial_{\mu}(\partial^{\mu
}\phi_{L})\phi_{R}=0 \label{64}%
\end{equation}
Now, suppose that $\phi_{L}$ is such that
\begin{equation}
\phi_{L}\tilde{\phi}_{L}=\phi_{L}\phi_{R}\neq0. \label{65}%
\end{equation}
Then $\phi_{L}$ is invertible and can be written
\begin{equation}
\phi_{L}=R\text{ }e^{\frac{1}{2}\gamma_{5}\beta}U \label{66}%
\end{equation}
where $R$,$\beta$ are scalar functions\footnote{The function $\beta$ is called
the Takabayshi angle.} (0-form fields) and for each $x\in M$, $U\in
\mathrm{Spin}_{1,3}^{0}(\simeq\mathrm{Sl}(2,\mathbb{C}))\subset\mathbb{R}%
_{1.3}^{0}$ and so $U\tilde{U}=\tilde{U}U=1.$ Putting $\rho=R^{2}$ from
Eqs.(\ref{56}) and (\ref{57}) we immediately get
\begin{align}
-\partial^{\mu}\phi_{L}  &  =[\mathbf{P}^{\mu}-\mathbf{W}^{\mu}]\phi
_{L}e^{-\gamma_{5}\beta}\gamma_{012},\label{67}\\
\partial^{\mu}\phi_{R}  &  =\gamma_{012}\phi_{R}e^{-\gamma_{5}\beta
}[\mathbf{P}^{\mu}+\mathbf{W}^{\mu}]. \label{68}%
\end{align}

\begin{remark}
Now, comes an important and crucial observation. Authors of
\emph{\cite{hc1,hc2}} write \emph{Eqs.(\ref{67}}) and \emph{(\ref{68})}
\ without the term $e^{-\gamma_{5}\beta}$. This means that they took the
Takabayashi angle as null. In \cite{rw2016a} we called \ invertible spinor
fields for which $\beta=0$ or $\pi$ classical spinor fields. The reason for
that name was that in \cite{rw2016a} we prove that the classical
Hamilton-Jacobi equation for a an electrical particle moving in an external
field $A$ is equivalent to a DHE satisfied by a classical Dirac-Hestenes
spinor field and that a DHE satisfied by a classical Dirac-Hestenes spinor
field is equivalent to the classical Hamilton-Jacobi equation. This classical
Dirac-Hestenes spinor field has the form
\begin{align}
\phi_{L}  &  =R\mathfrak{R}(\Pi)e^{S\gamma^{21}},\nonumber\\
\mathfrak{R}(\Pi)  &  =\frac{m+(\Pi+eA)\gamma^{0}}{\left[  2\left(  m+\Pi
_{0}+A_{0}\right)  \right]  ^{1/2}}, \label{69}%
\end{align}
with%
\begin{equation}
\Pi:-\partial S=mV+eA \label{70}%
\end{equation}
being the canonical momentum and
\begin{equation}
V=\phi_{L}\gamma^{0}\phi_{L}^{-1} \label{71}%
\end{equation}
being the \emph{(}$1$-form\emph{)} velocity field such that for each possible
world line of the classical particle \emph{(}parametrized with the proper
time\emph{)} $\sigma:\mathbb{R\rightarrow}M$ we have
\begin{equation}
\left.  \mathtt{\eta}(V,~)\right\vert _{\sigma}=\sigma_{\ast}=v \label{76}%
\end{equation}
where $v$ is the velocity of the particle.

To end this remark we present\emph{ \cite{rw2016a}} the correct generalized
Hamilton-Jacobi equation \emph{(GHJE)} that follows almost trivially in our
formalism for the \emph{DHE} satisfied by a general free \emph{DHSF} that we
write as%
\begin{equation}
\boldsymbol{\psi}=\boldsymbol{\rho}^{1/2}\mathcal{R}(\Pi)e^{\frac{\beta
\gamma^{5}}{2}}e^{S\gamma^{21}}=\boldsymbol{\psi}_{0}e^{\frac{\beta\gamma^{5}%
}{2}}e^{S\gamma^{21}}=e^{\frac{\beta\gamma^{5}}{2}}\boldsymbol{\psi}%
_{0}e^{S\gamma^{21}}.
\end{equation}
in interaction with an external electromagnetic field. It is, with
\emph{(}$\Pi=-\boldsymbol{\partial}S$ and $\boldsymbol{\psi}\gamma
^{0}\boldsymbol{\psi}^{-1}=e^{\beta\gamma^{5}}V$)%
\begin{equation}
-\boldsymbol{\partial}S=m\cos\beta V+eA+\langle m\sin\beta\gamma
^{5}V+(\boldsymbol{\partial}\ln\boldsymbol{\psi}_{0})\boldsymbol{\psi}%
\gamma^{21}\boldsymbol{\psi}^{-1}+\frac{1}{2}\gamma^{5}\boldsymbol{\partial
}(\ln\beta)\boldsymbol{\psi\rangle}_{1} \label{76aa}%
\end{equation}
where the following constraint must hold:%
\begin{equation}
\langle e\sin\beta\gamma^{5}V+(\boldsymbol{\partial}\ln\boldsymbol{\psi}%
_{0})\boldsymbol{\psi}\gamma^{21}\boldsymbol{\psi}^{-1}+\frac{1}{2}\gamma
^{5}\boldsymbol{\partial}(\ln\beta)\boldsymbol{\psi\rangle}_{3}=0 \label{76ac}%
\end{equation}

Thus the true "quantum potential" is
\begin{equation}
Q=\langle m\sin\beta\gamma^{5}V+(\boldsymbol{\partial}\ln\boldsymbol{\psi}%
_{0})\boldsymbol{\psi}\gamma^{21}\boldsymbol{\psi}^{-1}+\frac{1}{2}\gamma
^{5}\boldsymbol{\partial}(\ln\beta)\boldsymbol{\psi\rangle}_{1}\label{76ad}%
\end{equation}
which differs considerably form the usual Bohm quantum potential. Moreover and
contrary to the usual presentations of the de Broglie-Bohm theory the mass
parameter of the\ particle in the \emph{GHJE} \emph{(Eq.(\ref{76aa})) }is not
a constant, instead it is%
\begin{equation}
m^{\prime}=m\cos\beta.\label{76ae}%
\end{equation}

Some analogous results using classical like equations of motion isntead of the
Hamilton Jacobi like equation as above has been obtained in a series of
remarkable papers by Hestenes \cite{hestenes1,hestenes2,hestenes3}.
\end{remark}

Now, we returning to the analysis of papers \cite{hc1,hc2} where as shown
above used $\beta=0$. Under these conditions if we substitute Eqs.(\ref{67})
and (\ref{68}) in Eq.(\ref{63}) we get%

\begin{align}
&  \rho(\mathbf{P}^{\mu}\mathbf{P}_{\mu}+\mathbf{W}^{\mu}\mathbf{W}_{\mu
})\nonumber\\
&  +\rho(\partial_{\mu}\mathbf{P}^{\mu}\mathbf{J}-\mathbf{J}\partial_{\mu
}\mathbf{P}^{\mu})\nonumber\\
&  -\rho(\partial_{\mu}\mathbf{W}^{\mu}\mathbf{J}+\mathbf{J}\partial_{\mu
}\mathbf{W}^{\mu})-m^{2}\rho=0 \label{76a}%
\end{align}

Now, since $\langle\mathbf{W}^{\mu}\mathbf{W}_{\mu}\rangle_{2}=-\langle
\mathbf{W}^{\mu}\mathbf{W}_{\mu}\rangle_{2}=0$ we have that%
\begin{align}
&  (\mathbf{P}^{\mu}\mathbf{P}_{\mu}+\mathbf{W}^{\mu}\mathbf{W}_{\mu
})\nonumber\\
&  =\mathbf{P}^{\mu}\lrcorner\mathbf{P}_{\mu}+\langle\mathbf{W}^{\mu
}\mathbf{W}_{\mu}\rangle_{0}+\langle\mathbf{W}^{\mu}\mathbf{W}_{\mu}%
\rangle_{2}\nonumber\\
&  =\mathbf{P}^{\mu}\lrcorner\mathbf{P}_{\mu}+\mathbf{W}^{\mu}\lrcorner
\mathbf{W}_{\mu} \label{76c}%
\end{align}%
\begin{align}
&  (\partial_{\mu}\mathbf{P}^{\mu}\mathbf{J}-\mathbf{J}\partial_{\mu
}\mathbf{P}^{\mu})\nonumber\\
&  =2\partial_{\mu}\mathbf{P}^{\mu}\wedge\mathbf{J.} \label{76b}%
\end{align}
Since\footnote{Note that for any $A,B\in\sec%
{\textstyle\bigwedge\nolimits^{3}}
T^{\ast}M\hookrightarrow\sec\mathcal{C\ell}(M,\mathtt{\eta})$ always $\langle
AB\rangle_{4}=0$.} $\langle\partial_{\mu}\mathbf{W}^{\mu}\mathbf{J}\rangle
_{4}=0$ and $\langle\partial_{\mu}\mathbf{W}^{\mu}\mathbf{J}\rangle_{2}=-$
$\langle\mathbf{\tilde{J}}\partial_{\mu}\mathbf{\tilde{W}}^{\mu}\rangle
_{2}=-\langle\mathbf{J}\partial_{\mu}\mathbf{W}^{\mu}\rangle_{2}$ it is
\begin{equation}
-(\partial_{\mu}\mathbf{W}^{\mu}\mathbf{J}+\mathbf{J}\partial_{\mu}%
\mathbf{W}^{\mu})=-2\partial_{\mu}\mathbf{W}^{\mu}\lrcorner\mathbf{J.}
\label{76C}%
\end{equation}

Then, Eq.(\ref{76a}) becomes%

\begin{equation}
\mathbf{P}^{\mu}\lrcorner\mathbf{P}_{\mu}+\mathbf{W}^{\mu}\lrcorner
\mathbf{W}_{\mu}+2\partial_{\mu}\mathbf{P}^{\mu}\wedge\mathbf{J}%
-2\langle\partial_{\mu}\mathbf{W}^{\mu}\mathbf{J}\rangle_{0}-m^{2}=0
\label{76HC}%
\end{equation}
which implies that:%

\begin{gather}
\mathbf{P}^{\mu}\lrcorner\mathbf{P}_{\mu}+\mathbf{W}^{\mu}\lrcorner
\mathbf{W}_{\mu}-2\langle\partial_{\mu}\mathbf{W}^{\mu}\mathbf{J}\rangle
_{0}-m^{2}=0,\label{76e}\\
2\partial_{\mu}\mathbf{P}^{\mu}\wedge\mathbf{J=0.} \label{76f}%
\end{gather}

Now (with $\beta=0$), substituting Eqs.(\ref{67}) and (\ref{68}) in
Eq.(\ref{64}) we get%

\begin{gather}
\underset{X}{\underbrace{2\rho\mathbf{J}\partial^{\mu}\mathbf{P}_{\mu}%
+2\rho\mathbf{J}\partial^{\mu}\mathbf{W}_{\mu}+2\rho\partial^{\mu}%
\mathbf{P}_{\mu}\mathbf{J}-2\rho\partial^{\mu}\mathbf{W}_{\mu}\mathbf{J}}%
}\nonumber\\
+\text{ }\underset{Y}{\underbrace{\phi_{L}\gamma_{012}\boldsymbol{\partial
}^{\mu}\phi_{R}\mathbf{P}_{\mu}+\mathbf{P}_{\mu}\boldsymbol{\partial}^{\mu
}\phi_{L}\gamma_{012}\phi_{R}}}\nonumber\\
+\text{ }\underset{Z}{\underbrace{\phi_{L}\gamma_{012}\boldsymbol{\partial
}^{\mu}\phi_{R}\mathbf{W}_{\mu}+\mathbf{W}_{\mu}\boldsymbol{\partial}^{\mu
}\phi_{L}\gamma_{012}\phi_{R}}}%
\begin{array}
[c]{c}%
\\
=0\\
\end{array}
\label{77}%
\end{gather}

Calling the contents of the three lines of Eq.(\ref{77}) respectively $X,Y,Z$
we have:%
\begin{align}
X  &  =4\rho\partial^{\mu}\mathbf{P}_{\mu}\lrcorner\mathbf{J}\boldsymbol{+}%
2\rho\langle\mathbf{J}\partial^{\mu}\mathbf{W}_{\mu}\rangle_{0}+2\rho
\langle\mathbf{J}\partial^{\mu}\mathbf{W}_{\mu}\rangle_{2}-2\rho
\langle\partial^{\mu}\mathbf{W}_{\mu}\mathbf{J}\rangle_{0}-2\rho
\langle\partial^{\mu}\mathbf{W}_{\mu}\mathbf{J}\rangle_{2}\nonumber\\
&  4\rho\partial^{\mu}\mathbf{P}_{\mu}\lrcorner\mathbf{J}+4\rho\langle
\mathbf{J}\partial^{\mu}\mathbf{W}_{\mu}\rangle_{2}, \label{78}%
\end{align}%
\begin{align}
Y  &  =-\rho(\mathbf{P}_{\mu}\mathbf{W}^{\mu}+\mathbf{W}^{\mu}\mathbf{P}_{\mu
})=\nonumber\\
&  =-\rho\lbrack\langle\mathbf{P}_{\mu}\mathbf{W}^{\mu}\rangle_{2}%
+\langle\mathbf{P}_{\mu}\mathbf{W}^{\mu}\rangle_{4}+\langle\mathbf{W}^{\mu
}\mathbf{P}_{\mu}\rangle_{2}+\langle\mathbf{P}_{\mu}\mathbf{W}^{\mu}%
\rangle_{4}]\nonumber\\
&  =-2\rho\langle\mathbf{P}_{\mu}\mathbf{W}^{\mu}\rangle_{2}=-2\rho
\mathbf{P}_{\mu}\lrcorner\mathbf{W}^{\mu}, \label{79}%
\end{align}%
\begin{equation}
Z=-\rho(\mathbf{P}_{\mu}\mathbf{W}^{\mu}+\mathbf{W}^{\mu}\mathbf{P}_{\mu});
\label{80}%
\end{equation}
So, Eq.(\ref{77}) becomes%

\begin{equation}
\partial^{\mu}\mathbf{P}_{\mu}\lrcorner\mathbf{J}+\langle\mathbf{J}%
\partial^{\mu}\mathbf{W}_{\mu}\rangle_{2}-\mathbf{P}_{\mu}\lrcorner
\mathbf{W}^{\mu}=0. \label{84}%
\end{equation}

\begin{remark}
Authors of \cite{hc1,hc2} did not obtain \emph{Eqs.(\ref{76e}), (\ref{76f})}
and \emph{(\ref{84})} and moreover they interpreted the scalar part in
Eq.(\ref{76HC}) as being the GHJE in Bohm formalism. If this was the case we
would identify $(\mathbf{W}^{\mu}\lrcorner\mathbf{W}_{\mu}-2\langle
\partial_{\mu}\mathbf{W}^{\mu}\mathbf{J}\rangle_{0})$ as the quantum potential.

However we claim that such an identification is misleading in view of the
results presented in \textit{\cite{rw2016a}} and also from the fact that
authors of \emph{\cite{hc1,hc2}} wrongly identified $2\rho\mathbf{P}^{\mu}$
with the components $T_{D}^{\mu0}$ of the energy-momentum of the Dirac field,
which is a nonsequitur since the objects $2\rho\mathbf{P}^{\mu}$ are $1$-forms
and objects $T_{D}^{\mu0}$ are scalars. Moreover, for Dirac theory\ as
\emph{Eq.(76ad) }shows the quantum potential is a $1$-form field, not a scalar
function.$.$
\end{remark}

\begin{remark}
As a final remark, we emphasize that \emph{Eq.(\ref{84})} does not describe
the development of the spin. The equation of motion of the spin is the one
given by \emph{Eq.(\ref{38})}.
\end{remark}

\section{Conclusions}

Using the Clifford and spin-Clifford bundle formalisms and some few results of
the theory of extensor calculus we derived the conservation laws for the
probability current, energy-momentum and total angular momentum extensors
resulting from the structure of DHE satisfied by a Dirac-Hestenes spinor field
(DHSF) in interaction with the electromagnetic field. It was shown in a quite
simple way that the energy-momentum extensor\footnote{Related to theTetrode
tensor of the Dirac spinor field.} of the DHSF is not symmetric and that its
antisymmetric part is the \emph{source }of the spin extensor field of the free
DHSF. In particular it was shown that these conservation laws implies that in
spacetime regions with topology permitting the existence of an electromagnetic
field potential $A\neq0$ but with $F=0$ the energy-momentum and total angular
momentum of the sole DHSF are not conserved and this fact clearly is the
origin of the Bohm-Aharonov effect. And here it must be emphasize that since
this result can be directly derived \ for the DHE\ satisfied by a classical
DHSF equivalent to the classical relativistic HJE satisfied by a (spinning)
charged particle it is not of \ primely of quantum nature, more precisely, one
can predict that the motion of spinning charged particles in a region \ with
topology permitting $A\neq0$ but with $F=0$ is different from the motion of
spinning charged particles in a region $A=0$ but with $F=0$.

We also briefly recalled results obtained in \cite{rw2016a} where it was shown
that (a): that the classical\ relativistic Hamilton-Jacobi equation (HJE)
\ for a charged electrical particle in interaction with the electromagnetic
field is equivalent to a DHE satisfied by a special class of DHSF that we call
\emph{classical }DHSF which are characterized by having the Takabayashi angle
equal to $0$ or $\pi$ and moreover that the DHE satisfied by a classical DHSF
is equivalent to the HJE; (b) The identification of the correct relativistic
quantum potential resulting from the DHE in a de Broglie-Bohm approach
\cite{bh,debroglie} leading to a HJE like equation for the motion of
(spinning) charged particle. And equipped with these results and the ones of
Section 2 we analyzed results of \cite{hc1,hc2} were authors though that they
have disclosed the relativistic quantum potential for the Dirac particle. We
show that the results of those papers are equivocated \ by explicitly showing
with detailed calculations were authors get mislead.

In view of the recent interest on experiments to verify if the trajectories of
particles predicted by de Broglie-Bohm theory (see, e.g.,
\cite{englertetal,mahleretal,savantetal}) and pertinent criticisms, like,
e.g., in \cite{ck1,ck2,jonesetal} we think that our results are worth to be
appreciated. In particular, it clear for us that any prediction of
trajectories even in the nonrelativistic limit must be done using the non
relativistic approximation to the DHE, for as shown long ago in an important
paper by Guther and Hestenes \cite{gh1975} the Schr\"{o}dinger equation which
may be derived for the DHE in fact describes not a spinless particle, but a
particle in a spin auto state.

Concerning also the trajectories problem it is important to discover if
particles follows the integral lines of the velocity field as defined in
\cite{rw2016a} or the integral lines of the vector field $\mathtt{\eta
}(\boldsymbol{T}_{D}^{0},)=T_{D}^{0\mu}\frac{\partial}{\partial\mathrm{x}%
^{\mu}}$ (which are the ones appearing in the conservation laws of
energy-momentum and total angular momentum).

\appendix{}

\section{Notation}

In this paper the arena where physical phenomena is supposed to take place is
the Minkowski spacetime structure $(M,\boldsymbol{\eta},D,\tau
_{\boldsymbol{\eta}},\uparrow)$ where the manifold $M\simeq\mathbb{R}^{4}$,
$\boldsymbol{\eta}\in\sec T_{0}^{2}M$ is Minkowski metric, $D$ is the
Levi-Civita connection of $\boldsymbol{\eta}$. Moreover $M$ is oriented by
$\tau_{\boldsymbol{\eta}}\in\sec%
{\textstyle\bigwedge\nolimits^{4}}
T^{\ast}M$ and time oriented\footnote{See details in \cite{rc2007}.} by
$\uparrow$. We introduce in $M$ global coordinates $\{\mathrm{x}^{\mu}\}$ in
Einstein-Lorentz-Poincar\'{e} gauge. We put
\begin{equation}
e_{\mu}:=\frac{\partial}{\partial\mathrm{x}^{\mu}}=\partial_{\mu}%
,~~~~~~\gamma^{\mu}:=d\mathrm{x}^{\mu} \label{A1}%
\end{equation}
which are respectively global basis for $TM$ and $T^{\ast}M$. Then we can
write $\boldsymbol{\eta=}\eta_{\mu\nu}\gamma^{\mu}\otimes\gamma^{\nu}$. We
also introduce the metric of the cotangent bundle, i.e., $\mathtt{\eta}%
=\eta^{\mu\nu}e_{\mu}\otimes e_{\nu}$ and define the reciprocal basis of the
basis $\{\gamma^{\mu}\}$ as being $\{\gamma_{\mu}\}$, with \ $\mathtt{\eta
}(\gamma_{\mu},\gamma^{\nu})=\delta_{\mu}^{\nu}$. We denote the Clifford
bundle of differential forms by $\mathcal{C\ell}(M,\mathtt{\eta}).$In what
follows the Clifford product of sections of $\mathcal{C\ell}(M,\mathtt{\eta})$
is denoted by juxtaposition of symbols. In particular we have the fundamental
relation%
\begin{equation}
\gamma^{\mu}\gamma^{\nu}+\gamma^{\mu}\gamma^{\nu}=2\eta^{\mu\nu}. \label{A2}%
\end{equation}

We use the notation $\mathcal{C}\in\sec\mathcal{C\ell}(M,\mathtt{\eta})$ to
denote a general section of the Clifford bundle and since the bundle of
differential forms $%
{\textstyle\bigwedge}
T^{\ast}M$ shares with $\mathcal{C\ell}(M,\mathtt{\eta})$ for each $x\in M$
the same vector space we can write that%
\begin{equation}
\mathcal{C=}%
{\textstyle\sum\nolimits_{i=0}^{4}}
\mathcal{C}_{i} \label{A22}%
\end{equation}
where $\mathcal{C}_{i}\in\sec%
{\textstyle\bigwedge\nolimits^{i}}
T^{\ast}M\hookrightarrow\sec\mathcal{C\ell}(M,\mathtt{\eta})$. We also use the
projection operators%
\begin{equation}
\langle\rangle_{i}:\mathcal{C}\in\sec\mathcal{C\ell}(M,\mathtt{\eta}%
)\mapsto\mathcal{C}_{i}\in\sec%
{\textstyle\bigwedge\nolimits^{i}}
T^{\ast}M\hookrightarrow\sec\mathcal{C\ell}(M,\mathtt{\eta}). \label{a22b}%
\end{equation}

We recall that if $\mathcal{A}_{r}\in\sec%
{\textstyle\bigwedge\nolimits^{r}}
T^{\ast}M\hookrightarrow\sec\mathcal{C\ell}(M,\mathtt{\eta}),\mathcal{B}%
_{s}\in\sec%
{\textstyle\bigwedge\nolimits^{s}}
T^{\ast}M\hookrightarrow\sec\mathcal{C\ell}(M,\mathtt{\eta})$%

\begin{equation}
\mathcal{A}_{r}\mathcal{B}_{s}=\langle\mathcal{A}_{r}\mathcal{B}_{s}%
\rangle_{\left\vert r-s\right\vert }+\langle\mathcal{A}_{r}\mathcal{B}%
_{s}\rangle_{\left\vert r-s\right\vert +2}+\cdots+\langle\mathcal{A}%
_{r}\mathcal{B}_{s}\rangle_{r+s}=\sum_{k=0}^{m}\langle\mathcal{A}%
_{r}\mathcal{B}_{s}\rangle_{|r-s|+2k}, \label{T54a}%
\end{equation}
Moreover, for $\mathcal{A},\mathcal{C}\in\sec\mathcal{C\ell}(M,\mathtt{\eta})$
the following identity that will be used several times in the calculations of
main text holds%
\begin{equation}
\langle\mathcal{AC}\rangle_{r}=(-1)^{r\frac{(r-1)}{2}}\langle\mathcal{\tilde
{B}\tilde{C}}\rangle_{r} \label{T54B}%
\end{equation}
where%
\begin{gather}
\overset{\symbol{126}}{}:\sec%
{\textstyle\bigwedge\nolimits^{r}}
T^{\ast}M\hookrightarrow\sec\mathcal{C\ell}(M,\mathtt{\eta})\mapsto\sec%
{\textstyle\bigwedge\nolimits^{r}}
T^{\ast}M\hookrightarrow\sec\mathcal{C\ell}(M,\mathtt{\eta}),\nonumber\\
\mathcal{C}_{r}\mapsto\mathcal{\tilde{C}}_{r}=(-1)^{r\frac{(r-1)}{2}}.
\label{t54c}%
\end{gather}
We have also used the fact that $\mathcal{B}_{r}\in\sec%
{\textstyle\bigwedge\nolimits^{r}}
T^{\ast}M\hookrightarrow\sec\mathcal{C\ell}(M,\mathtt{\eta})$, $\mathcal{B}%
_{s}\in\sec%
{\textstyle\bigwedge\nolimits^{s}}
T^{\ast}M\hookrightarrow\sec\mathcal{C\ell}(M,\mathtt{\eta})$ with $r\leq s$
the right and left contractions are related by%
\begin{equation}
\mathcal{B}_{r}\lrcorner\mathcal{C}_{s}=\langle\mathcal{B}_{r}\mathcal{C}%
_{s}\rangle_{\left\vert r-s\right\vert }=(-1)^{r\frac{(r-1)}{2}}%
\langle\mathcal{C}_{s}\mathcal{B}_{r}\rangle_{\left\vert r-s\right\vert
}=\mathcal{C}_{s}\llcorner\mathcal{B}_{r}. \label{T54E}%
\end{equation}
Also,%
\begin{equation}
\mathcal{B}_{r}\wedge\mathcal{C}_{s}=\langle\mathcal{B}_{r}\mathcal{C}%
_{s}\rangle_{\left\vert r-s\right\vert }=(-1)^{r}\mathcal{C}_{s}%
\wedge\mathcal{B}_{r}. \label{t54f}%
\end{equation}

\begin{remark}
We remark that with our convetions the scalar product \emph{(}denoted by the
symbol $\cdot$\emph{)} of multiforms $\mathcal{B}_{r},\mathcal{C}_{s}\in\sec%
{\textstyle\bigwedge\nolimits^{r}}
T^{\ast}M\hookrightarrow\sec\mathcal{C\ell}(M,\mathtt{\eta})$ is null for
$r\neq s$ and for $r=s$ it is related to the contractions by the following
relation%
\[
\mathcal{B}_{r}\lrcorner\mathcal{C}_{r}=\mathcal{B}_{r}\llcorner
\mathcal{C}_{r}=\mathcal{\tilde{B}}_{r}\cdot\mathcal{C}_{r}=\mathcal{B}%
_{r}\cdot\mathcal{\tilde{C}}_{r}%
\]

\end{remark}

\begin{remark}
Moreover,\ we wrote the volume element $\tau_{\mathtt{\eta}}\in\sec%
{\textstyle\bigwedge\nolimits^{4}}
T^{\ast}M\hookrightarrow\sec\mathcal{C\ell}(M,\mathtt{\eta})$ as%
\begin{equation}
\tau_{\mathtt{\eta}}=\gamma^{0}\wedge\gamma^{1}\wedge\gamma^{2}\wedge
\gamma^{3}=\gamma^{0}\gamma^{1}\gamma^{2}\gamma^{3}=:\gamma^{5}. \label{vol}%
\end{equation}

\end{remark}

Besides $\mathcal{C\ell}(M,\mathtt{\eta})$ we need also the so-called left and
right spin-Clifford bundles denoted $\mathcal{C\ell}_{\mathrm{Spin}_{1,3}^{e}%
}^{L}(M,\mathtt{\eta})$ and $\mathcal{C\ell}_{\mathrm{Spin}_{1,3}^{e}}%
^{R}(M,\mathtt{\eta})$ and the complexified left and right spin-Clifford
bundles denoted $\mathbb{C}\mathcal{\ell}_{\mathrm{Spin}_{1,3}^{e}}%
^{L}(M,\mathtt{\eta})$ and $\mathbb{C}\mathcal{\ell}_{\mathrm{Spin}_{1,3}^{e}%
}^{R}(M,\mathtt{\eta})$. These bundles are all trivial\footnote{Indeed, all
spinor bundles over a $4$-dimensional Lorentzian spacetime are trivial. See
(\cite{geroch1,geroch2})} and have the structures%
\begin{gather}
\mathcal{C\ell}(M,\mathtt{\eta})=P_{\mathrm{Spin}_{1,3}^{e}}(M)\times
_{\mathrm{Ad}}\mathbb{R}_{1,3},\nonumber\\
\mathcal{C\ell}_{\mathrm{Spin}_{1,3}^{e}}^{L}(M,\mathtt{\eta}%
)=P_{\mathrm{Spin}_{1,3}^{e}}(M)\times_{l}\mathbb{R}_{1,3},~~~\mathcal{C\ell
}_{\mathrm{Spin}_{1,3}^{e}}^{L}(M,\mathtt{\eta})=P_{\mathrm{Spin}_{1,3}^{e}%
}(M)\times_{r}\mathbb{R}_{1,3},\nonumber\\
\mathbb{C}\mathcal{\ell}_{\mathrm{Spin}_{1,3}^{e}}^{L}(M,\mathtt{\eta
})=P_{\mathrm{Spin}_{1,3}^{e}}(M)\times_{l}\mathbb{R\otimes R}_{1,3}%
,~~~\mathbb{C}\mathcal{\ell}_{\mathrm{Spin}_{1,3}^{e}}^{R}(M,\mathtt{\eta
})=P_{\mathrm{Spin}_{1,3}^{e}}(M)\times_{r}\mathbb{R\otimes R}_{1,3}.
\label{A3}%
\end{gather}
where $\mathbb{R}_{1,3}\simeq\mathbb{H(}2)$ is the spacetime algebra,
$P_{\mathrm{Spin}_{1,3}^{e}}(M)$ is a principal bundle called spinor structure
bundle. Moreover, \textrm{Ad}$:\mathrm{Spin}_{1,3}^{e}\rightarrow
\mathrm{Aut}(\mathbb{R}_{1,3})$ with \textrm{Ad}$_{u}X=uXu^{-1}$ and $l$ and
$r$ denotes respectively the representations of $\mathrm{Spin}_{1,3}%
^{e}(\simeq\mathrm{Sl}(2,\mathbb{C}))$ on $\mathbb{R}_{1,3}$ given
respectively by $l(u)X=uX$ and $r(u)X=Xu$. Left and right \emph{Dirac-Hestenes
spinor fields} (DHSF) are respectively sections of $\mathcal{C\ell
}_{\mathrm{Spin}_{1,3}^{e}}^{0L}(M,\mathtt{\eta})$ and $\mathcal{C\ell
}_{\mathrm{Spin}_{1,3}^{e}}^{0R}(M,\mathtt{\eta})$, the even subbundles of
$\mathcal{C\ell}_{\mathrm{Spin}_{1,3}^{e}}^{L}(M,\mathtt{\eta})$ and
$\mathcal{C\ell}_{\mathrm{Spin}_{1,3}^{e}}^{R}(M,\mathtt{\eta}).$We define
also the bundles $S^{L}(M)$ and $S^{R}(M)$ such that
\begin{equation}
S^{L}(M)=P_{\mathrm{Spin}_{1,3}^{e}}(M)\times_{l}I^{R},~~~S^{l}%
(M)=P_{\mathrm{Spin}_{1,3}^{e}}(M)\times_{r}I^{R} \label{A4}%
\end{equation}
with the minimal ideals $I^{L}$ and $I^{R}$ in $\mathbb{C\otimes}%
\mathbb{R}_{1,3}\simeq\mathbb{R}_{4,1}\simeq\mathbb{C(}4)$ being respectively%
\begin{equation}
I^{L}=(\mathbb{R\otimes R}_{1,3})f\mathbf{,~~~}I^{R}=\tilde{f}%
(\mathbb{R\otimes R}_{1,3}). \label{A5}%
\end{equation}
and generated by the idempotents $f$ and $\tilde{f}$ where
\begin{equation}
f=\frac{1}{2}(1+\gamma_{0})\frac{1}{2}(1+\mathrm{i}\gamma_{1}\gamma_{2})\in
I\subset\mathbb{C\otimes}\mathbb{R}_{1,3},\text{ \quad}\mathrm{i}=\sqrt{-1}.
\label{D13a}%
\end{equation}

The sections of $S^{l}(M)$ \ can be represented in $\mathbb{C}(4)$ and there
is a $1-1$ correspondence with\emph{\ ideal sections }of the bundle
$\mathcal{S}^{l}(M)=P_{\mathrm{Spin}_{1,3}^{e}}(M)\times_{l}\mathbb{C}%
(4)\mathbf{f}$ where $\mathbf{f}$, the representation of $f$ in $\mathbb{C(}%
4)$ is given by
\begin{equation}
\mathbf{f}=\frac{1}{2}(1+\underline{\gamma}_{0})\frac{1}{2}(1+\mathrm{i}%
\underline{\gamma}_{1}\underline{\gamma}_{2}),\text{ \quad}\mathrm{i}%
=\sqrt{-1}. \label{D14}%
\end{equation}
and where
\[
\mathcal{\gamma}_{0}\mapsto\underline{\gamma}_{0}=\left(
\begin{array}
[c]{cc}%
\mathbf{1}_{2} & 0\\
0 & \mathbf{-1}_{2}%
\end{array}
\right)  ;~~~~~\mathcal{\gamma}_{i}\mapsto\underline{\gamma}_{i}=\left(
\begin{array}
[c]{cc}%
0 & -\sigma_{i}\\
\sigma_{i} & 0
\end{array}
\right)  ,
\]
where $\mathbf{1}_{2}$ is the unit $2\times2$ matrix and $\sigma_{i}$,
($i=1,2,3$) are the standard \textit{Pauli matrices}.

Finally, we recall that:

(i) The\ ideal sections of $\mathcal{S}^{l}(M)$ are\ $1$-$1$ correspondence
with\ the sections of the bundle the Dirac spinor fields represented as
columns complex vectors in the standard formalism used in Physics textbooks
and which are sections of the bundle%
\begin{equation}
\mathcal{S}^{l}(M)=P_{\mathrm{Spin}_{1,3}^{e}}(M)\times_{l}\mathbb{C}^{4}.
\label{A6}%
\end{equation}
(ii) Sections of the bundles $\mathcal{C\ell}_{\mathrm{Spin}_{1,3}^{e}}%
^{0L}(M,\mathtt{\eta})$ and $\mathcal{C\ell}_{\mathrm{Spin}_{1,3}^{e}}%
^{0R}(M,\mathtt{\eta})$ are represented by some well defined equivalence
classes of sections of the even subbundle $\mathcal{C\ell}^{0}(M,\mathtt{\eta
})$ of $\mathcal{C\ell}(M,\mathtt{\eta})$. A representative is given once we
fix a spinorial frame.

In what follows we suppose that a spinorial frame has been fixed in order to
simplify our notation. Details are given in \cite{r2004,mr2004,rc2007}. Here
we recall that the matrix representation in $\mathbb{C(}4)$ of $\Phi_{L}\in
I^{R}$ will be denoted by the same letter in boldface, i.e., $\Phi_{L}%
\mapsto\mathbf{\Phi}_{L}\in\mathbf{I}=\mathbb{C}(4)\mathbf{f}$. We have (see,
e.g., \cite{fro,lounesto,vr2016}
\begin{equation}
\mathbf{\Phi}_{L}=\left(
\begin{array}
[c]{llll}%
\psi_{1} & 0 & 0 & 0\\
\psi_{2} & 0 & 0 & 0\\
\psi_{3} & 0 & 0 & 0\\
\psi_{4} & 0 & 0 & 0
\end{array}
\right)  ,\text{ \ \ \ }\psi_{i}\in\mathbb{C}. \label{D15}%
\end{equation}
Any element $\Phi_{L}\in I$ can be written
\begin{equation}
\Phi_{L}=\phi_{L}f,~~~~~\phi_{L}\in\mathbb{R}_{1,3}^{0}\subset\mathbb{C\otimes
}\mathbb{R}_{1,3} \label{D16}%
\end{equation}
The matrix representation of $\phi_{L}$ (a representative of a
\emph{Dirac-Hestenes spinor }in a given spin frame) in $\mathbb{C(}4)$ will be
denoted by the same letter in boldsymbol, i.e.:%

\begin{equation}
\boldsymbol{\phi}_{L}=\left(
\begin{array}
[c]{cccc}%
\psi_{1} & -\psi_{2}^{\ast} & \psi_{3} & \psi_{4}^{\ast}\\
\psi_{2} & \psi_{1}^{\ast} & \psi_{4} & -\psi_{3}^{\ast}\\
\psi_{3} & \psi_{4}^{\ast} & \psi_{1} & -\psi_{2}^{\ast}\\
\psi_{4} & -\psi_{3}^{\ast} & \psi_{2} & \psi_{1}^{\ast}%
\end{array}
\right)  \label{D17}%
\end{equation}

\begin{remark}
When \emph{(}global\emph{)} coordinates in Einstein-Lorentz-Poincar\'{e} gauge
are introduced in $M$ the spin-Dirac operator $\partial^{s}$\ acting on
sections of the left spinor bundle will be denoted simply by $\partial
=\underline{\gamma}^{\mu}\partial_{\mu}$. In this case, taking moreover into
account that all spinor bundles in Lorentzian spacetime structures are trivial
\cite{geroch1,geroch2} we use in what follows a convenient obvious notation
writing the Dirac equation satisfied by a \textit{spinor field} $\mathbf{\Phi
}_{L}:M\simeq\mathbb{R}^{4}\rightarrow\mathbf{I}=\mathbb{C}(4)\mathbf{f}$ as%
\begin{equation}
\mathrm{i}\partial\mathbf{\Phi}_{L}-m\mathbf{\Phi}_{L}=0 \label{D18}%
\end{equation}

\end{remark}

In this paper we denote the Dirac conjugate\footnote{Note that the Dirac
conjugate spinor of $\mathbf{\Phi}_{L}$ is usually denote by $\mathbf{\bar
{\Phi}}_{L}:$ $=\mathbf{\Phi}_{L}^{\dagger}\underline{\gamma}_{0}$. We did not
use this notation here because as already recalled we use in \cite{rc2007}
then the symbol $\overline{}$ to denote the main involution operator} of a
\emph{spinor field\ }$\mathbf{\Phi}_{L}$ \emph{as}
\begin{equation}
\mathbf{\Phi}_{R}:=\mathbf{\Phi}_{L}^{\dagger}\underline{\gamma}_{0}
\label{D19}%
\end{equation}
we can easily show (as it is well known) that it satisfies the following Dirac
conjugate equation%
\begin{equation}
\mathrm{i}\mathbf{\Phi}_{R}\overleftarrow{\partial}+m\mathbf{\Phi}%
_{R}=0,~~~~\mathbf{\Phi}_{R}\overleftarrow{\partial}:=\partial_{\mu
}\mathbf{\Phi}_{R}\underline{\gamma}^{\mu}. \label{D20}%
\end{equation}

\subsection{The Dirac-Hestenes Equation}

\begin{remark}
Now, we recall that the field $\phi_{L}:M\simeq\mathbb{R}^{4}\rightarrow
\mathbb{R}_{1,3}^{0}\subset\mathbb{C\otimes}\mathbb{R}_{1,3}$ which as already
said above is a \emph{representative} of a Dirac-Hestenes spinor field
\emph{(}an object which is a section of a \emph{Spin-Clifford bundle) in }the
\emph{Clifford bundle }once a spin frame is fixed. To have in mind that
$\phi_{L}$ is a representative of a DHSF is extremely important in order to
avoid misconceptions. So, take notice that \emph{(}using coordinates in
Einstein-Lorentz-Poincar\'{e} gauge for $M$\emph{) }the representative\ in the
Clifford bundle of the spin-Dirac operator $\boldsymbol{\partial}^{s}$\ acting
on the bundle of Dirac-Hestenes spinor fields\emph{\footnote{See details in
Chapter 7 of \cite{rc2007}.}} will be simply denoted by%
\begin{equation}
\boldsymbol{\partial}=\gamma^{\mu}\partial_{\mu}. \label{d20a}%
\end{equation}

\end{remark}

Under these conditions one can easily shown that the Dirac equation
\emph{(\ref{D18})}is equivalent to the following equation which is known as
the Dirac-Hestenes equation%
\begin{equation}
\boldsymbol{\partial}\phi_{L}\gamma_{21}-m\phi_{L}\gamma_{0}=0, \label{D21}%
\end{equation}

To continue we observe that if we multiply this equation on the left by the
idempotent $f$ (Eq.(\ref{D13a})) we immediately get recalling Eq.(\ref{D16})
satisfies the following differential equation.%
\begin{equation}
\mathrm{i}\boldsymbol{\partial}\Phi_{L}-m\Phi_{L}=0, \label{D22}%
\end{equation}
whose matrix representation in $\mathbb{C}^{4}$ is of course given by
Eq.(\ref{D18}).

The object representing $\mathbf{\Phi}_{R}$ in the Clifford bundle formalism
is \cite{lounesto}
\[
\Phi_{R}:=\tilde{\Phi}_{L}^{\ast}:M\simeq\mathbb{R}^{4}\rightarrow
I_{R}=\tilde{f}^{\ast}(\mathbb{C\otimes}\mathbb{R}_{1,3})
\]
where the idempotent
\[
\tilde{f}^{\ast}=\frac{1}{2}(1-i\gamma_{21})\frac{1}{2}(1+\gamma_{0})
\]
generates (for each $x\in M$) the ideal $I_{R}=\tilde{f}^{\ast}%
(\mathbb{C\otimes}\mathbb{R}_{1,3})$. Of course,
\begin{equation}
\Phi_{R}:=\tilde{\Phi}_{L}^{\ast}=\frac{1}{2}(1-i\gamma_{21})\frac{1}%
{2}(1+\gamma_{0})\tilde{f}\text{ }\tilde{\phi}_{L},~~~~~~\phi_{R}:=\tilde
{\phi}_{L} \label{D23}%
\end{equation}
and we immediately get
\begin{equation}
\mathrm{i}\Phi_{R}\overleftarrow{\boldsymbol{\partial}}+m\Phi_{R}=0.
\label{D25}%
\end{equation}

\end{document}